\documentclass[11pt,reqno]{amsart}

\usepackage{amscd,amssymb,amsmath,amsthm}
\usepackage[arrow,matrix]{xy}
\usepackage{graphicx}
\usepackage{epstopdf}
\usepackage{color}
\usepackage{cite}
\numberwithin{equation}{section} 

\begin{document}
\title[Three state hard core models]{
Periodic Gibbs measures for three-state hard-core models in the case Wand
}

\author{R. M. Khakimov, K.O.Umirzakova}

\address{R.\ M.\ Khakimov \\ Institute of Mathematics, Namangan State University,
316, Uychi str., 160136, Namangan, Uzbekistan.}

\address{K.\ O.\ Umirzakova\\ Namangan State University,
316, Uychi str., 160136, Namangan, Uzbekistan.}
\email {rustam-7102@rambler.ru \ \ kamola-0983@mail.ru}

\begin{abstract} We consider fertile three-state Hard-Core (HC) models with the activity parameter $\lambda>0$
on a Cayley tree. It is known that there exist four types of such models: wrench, wand, hinge, and pipe. These
models arise as simple examples of loss networks with nearest-neighbor exclusion. In the case wand on a Cayley
tree of order $k\geq2$, exact critical values $\lambda>0$ are found for which two-periodic
Gibbs measures are not unique. Moreover, we study the extremality of the
existing two-periodic Gibbs measures on a Cayley tree of order two.
\end{abstract}
\maketitle



\section{Introduction}

Description of all limit Gibbs measures for a given
Hamiltonian is one of the main problems of the theory of Gibbs measures.
It is known that each Gibbs measure is associated with one phase of the physical system.
Therefore, in the theory of Gibbs measures, one of the important problems is the existence
of a phase transition, i.e., when the physical system changes its state when the temperature changes.
This occurs when the Gibbs measure is not unique. In this case, the temperature at which the state
of the physical system changes is usually called the critical temperature.
Moreover, it is known that for continuous Hamiltonians (see \cite {FV}) it is known that the Gibbs measures form a non-empty convex compact set in the space of all probability measures endowed with the weak topology (see, e.g., \cite[Chapter 7]{6}). The set of the Gibbs measures on $\mathbb Z^d$   is the convex hull of the set of all limit Gibbs measures (See \cite {C}).

In this connection, it is particularly interesting to describe all the extreme points of this convex set, i.e., the extreme Gibbs measures.

The definition of the Gibbs measure and other concepts related to Gibbs measure theory can be found,
for example, in \cite {6}, \cite {Pr}, \cite {R}, \cite {Si}. Although there are many works devoted to studying Gibbs measures, a complete description of all limit Gibbs measures has not yet been obtained for any of the models on Cayley trees.

Hard constraints arise in fields as diverse as combinatorics, statistical mechanics, and
telecommunications. In particular, the hard-core model arises in the study of random independent
sets of a graph \cite {bw1}, \cite {gk}, the study of gas molecules on a lattice \cite {B}, and in
the analysis of multi-casting in telecommunication networks \cite {kf}, \cite {mrs}.

Mazel and Suhov introduced and studied the HC model on the $d$-dimensional lattice $ \mathbb Z ^ d$ \cite {Maz}.
In \cite{bw}, fertile HC models were identified that correspond to graphs of the hinge, pipe, wand
and wrench types. The Gibbs measures for HC models with three states on the Cayley tree of order $k\geq1$ were
studied in \cite {bw}, \cite {XR1},  \cite {MRS}, \cite {RKh1}, \cite {Ro},   \cite {7}. In particular, in \cite{Ro} and \cite {XR1} in the "wand"  case,
a full description of translation-invariant Gibbs measures (TIGM) is given on the Cayley tree of order two and three, respectively.
Also in this case, the existence of at least three TIMGs on a Cayley tree of arbitrary order is proved in \cite {RKh1}. Moreover,
in \cite {RKh1} areas of the (non) extremality of TIMG on the Cayley tree of order $k=2$ were found.
Work \cite {KKR} is devoted to the study of translation-invariant and periodic Gibbs measures for
three-state HC models with an external field. Translation-invariant and periodic Gibbs measures in "hinge", "pipe" and "wrench" cases were studied in \cite {XR1},  \cite {MRS}, \cite {RKh1}, \cite {Ro}. In the "wand"  case, periodic measures have not yet been studied. See Chap. 7 in \cite {R} for other HC model properties and their generalizations on a Cayley tree.

In this paper, we study periodic Gibbs measures for a fertile three-state HC model in the case of a "wand" on a homogeneous Cayley tree.
In this case on a Cayley tree of arbitrary order under certain conditions, the translation invariance of the $G^{(2)}_k$-periodic Gibbs measures is proved. In addition on the Cayley tree of orders two and three under certain conditions an exact critical value $\lambda_{cr}$ is found such that, for $\lambda\geq\lambda_{cr}$ there exists exactly one $G^{(2)}_k$-periodic Gibbs measure, which is translation-invariant and for $0<\lambda<\lambda_{cr}$ there are exactly three $G^{(2)}_k$-periodic Gibbs measures, one of which is translation-invariant, and the other two are $G^{(2)}_k$-periodic (non translation-invariant). Also under certain conditions, we find explicit value $\lambda_{cr}(k)$ such that for $0<\lambda<\lambda_{cr}$ there exist no less than two $G^{(2)}_k$-periodic (non translation-invariant) Gibbs measures on a Cayley tree of order $k\geq2$. Moreover, we check extremality of the $G^{(2)}_k$-periodic Gibbs measures existing on the Cayley tree of order two.

\section{Preliminaries}\

The Cayley tree $\Im^k$
of order $ k\geq 1 $ is an infinite tree,
i.e., a connected graph without cycles, such that exactly $k+1$ edges
originate from each vertex. Let $\Im^k=(V,L,i)$, where $V$ is the
set of vertices $\Im^k$, $L$ is the set of edges and $i$ is the
incidence function setting each edge $l\in L$ into correspondence
with its endpoints $x, y \in V$. If $i (l) = \{ x, y \} $, then
the vertices $x$ and $y$ are called the {\it nearest neighbors},
denoted by $l = \langle x, y \rangle $.

For a fixed point $x^0\in V$,
$$W_n=\{x\in V\,| \, d(x,x^0)=n\}, \qquad V_n=\bigcup_{m=0}^n W_m, \qquad L_n=\{\langle x,y\rangle\in L| \, x,y\in V_n\},$$
where $d(x,y)$ is the distance between
vertices $x$ and $y$ on a Cayley tree, i.e.,
the number of edges of the shortest path connecting  $x$ and $y$.

Write  $x\prec y$, if the path from  $x^0$ to $y$ goes through $x$.
Call vertex $y$ a direct successor of $x$ if $y\succ x$ and $x,y$
are nearest neighbors. Note that in $\Im^k$ any vertex $x\neq x^0$
has $k$ direct successors and $x^0$ has $k+1$ direct successors. Denote by
$S(x)$ the set of direct successors of $x$, i.e. if $x\in W_n$, than
$$S(x)=\{y_i\in W_{n+1} |  d(x,y_i)=1, i=1,2,\ldots, k \}.$$

\emph{HC model.} Let $\Phi = \{0,1,2\}$ and $\sigma\in \Omega=\Phi^V$ be a configuration on $V$,
i.e., $\sigma=\{\sigma(x)\in \Phi: x\in V\}$. In this model, each vertex $x$
is assigned one of the values $\sigma (x)\in \Phi=\{0,1,2\}$.
The values $\sigma (x)=1,2$ mean that the
vertex $x$ is `occupied', and $\sigma (x)=0$ means that $x$ is `vacant'.
We let $\Omega$ denote the set of all configurations on $V$. Configurations in
$V_n$ and $W_n$ can be defined similarly, with the set of all
configurations in $V_n$ and $W_n$ denoted by $\Omega_{V_n}$ and $\Omega_{W_n}$.

We consider the set $\Phi$ as the set of vertices of a graph $G$.
We use the graph $G$  to define a $G$-admissible
configuration as follows. A configuration $\sigma$ is called a
$G$-\textit{admissible configuration} on the Cayley tree (in $V_n$ or
in $W_n$), if $\{\sigma (x),\sigma (y)\}$ is the edge of the graph $G$
for any pair of nearest neighbors $x,y$ in $V$ (in $V_n$). We
let $\Omega^G$ ($\Omega_{V_n}^G$) denote the set of $G$-admissible configurations.

The activity set \cite{bw} for a graph $G$ is a function $\lambda:G
\to R_+$ from the set $G$  to the set of positive real
numbers. The value $\lambda_i$ of the function $\lambda$ at the vertex
$i\in\{0,1,2\}$ is called the vertex activity.

For given $G$ and $\lambda$ we define the Hamiltonian of the $G-$HC model as
 $$H^{\lambda}_{G}(\sigma)=\left\{%
\begin{array}{ll}
     \sum\limits_{x\in{V}}{\log \lambda_{\sigma(x)},} \ \ \ $ if $ \sigma \in\Omega^{G} $,$ \\
   +\infty ,\ \ \ \ \ \ \ \ \ \  \ \ \ $  \ if $ \sigma \ \notin \Omega^{G} $.$ \\
\end{array}%
\right. $$

The union of configurations $\sigma_{n-1}\in\Phi ^ {V_{n-1}}$ and $\omega_n\in\Phi
^ {W_{n}}$ is determined by the following formula:
$$
\sigma_{n-1}\vee\omega_n=\{\{\sigma_{n-1}(x), x\in V_{n-1}\},
\{\omega_n(y), y\in W_n\}\}.
$$

Let $\mathbf{B}$ be the $\sigma$-algebra generated by cylindric subsets of $\Omega^{G}.$ For any arbitrary $n$ we let $\mathbf{B}_{V_n}=\{\sigma\in\Omega^{G}:
\sigma|_{V_n}=\sigma_n\}$, where $\sigma|_{V_n}$ is the restriction of $\sigma$ to $V_n$ and $\sigma_n: x\in V_n
\mapsto \sigma_n(x)$ is an admissible configuration in $V_n$, denote subalgebra of $\mathbf{B}.$

\textbf{Definition 1}. For $\lambda >0$ the HC model Gibbs measure is a probability measure $\mu$ on $(\Omega^{G}, \textbf{B})$ such that for any $n$ and $\sigma_n\in \Omega_{V_n}^{G}$, we have
$$
\mu \{\sigma \in \Omega^{G}:\sigma|_{V_n}=\sigma_n\}=
\int_{\Omega^{G}}\mu(d\omega)P_n(\sigma_n|\omega_{W_{n+1}}),
$$
where
$$
P_n(\sigma_n|\omega_{W_{n+1}})=\frac{e^{-H^{\lambda}_{G}(\sigma_n)}}{Z_{n}
(\lambda ; \omega |_{W_{n+1}})}\textbf{1}(\sigma_n \vee \omega
|_{W_{n+1}}\in\Omega_{V_{n+1}}^{G}).
$$

Here $Z_n (\lambda ; \omega|_{W_{n+1}})$ is the normalization multiplier with the boundary condition $\omega|_{W_{n+1}}$:
$$
Z_n (\lambda ; \omega|_{W_{n+1}})=\sum_{\widetilde{\sigma}_n \in
\Omega_{V_n}}
e^{-H^{\lambda}_{G}(\widetilde{\sigma}_n)}\textbf{1}(\widetilde{\sigma}_n\vee
\omega|_{W_{n+1}}\in \Omega_{V_{n+1}}^{G}).
$$

\textbf{Definition 2.}\cite{bw} A graph is said to be fertile if there is a set
of activities $\lambda$ such that the corresponding Hamiltonian has at least two
translation-invariant Gibbs measures.

In this paper we consider the case $\lambda_0=1, \
\lambda_1=\lambda_2=\lambda \ $ and we study periodic Gibbs measures in the case fertile graph
$G=\textit{wand}$:
\[
\begin{array}{ll}
\mbox{\it wand}: &  \{0,1\}\{0,2\}\{1,1\}\{2,2\}.\\
\end{array} \]

For $\sigma_n\in\Omega_{V_n}^G$ we let
$$\#\sigma_n=\sum\limits_{x\in V_n}{\mathbf 1}(\sigma_n(x)\geq 1)$$
denote the number of occupied vertices in $\sigma_n$.

Let $z:\;x\mapsto z_x=(z_{0,x}, z_{1,x}, z_{2,x}) \in R^3_+$
be a vector-valued function on $V$. For $n=1,2,\ldots$ and $\lambda>0$,
we consider the probability measure $\mu^{(n)}$ on $\Omega_{V_n}^G$
defined as
\begin{equation}\label{rus2.1}
\mu^{(n)}(\sigma_n)=\frac{1}{Z_n}\lambda^{\#\sigma_n} \prod_{x\in
W_n}z_{\sigma(x),x},
\end{equation}
where $Z_n$ is a normalization factor,
$$
Z_n=\sum_{{\widetilde\sigma}_n\in\Omega^G_{V_n}}
\lambda^{\#{\widetilde\sigma}_n}\prod_{x\in W_n}
z_{{\widetilde\sigma}(x),x}.
$$

The probabilistic measure $\mu^{(n)}$ is said to be consistent if for all $n\geq 1$ and
any $\sigma_{n-1}\in\Omega^G_{V_{n-1}}$:

\begin{equation}\label{rus2.2}
\sum_{\omega_n\in\Omega_{W_n}}
\mu^{(n)}(\sigma_{n-1}\vee\omega_n){\mathbf 1}(
\sigma_{n-1}\vee\omega_n\in\Omega^G_{V_n})=
\mu^{(n-1)}(\sigma_{n-1}).
\end{equation}

In this case, there is a unique measure $\mu$ on $(\Omega^G,
\textbf{B})$ such that
$$\mu(\{\sigma|_{V_n}=\sigma_n\})=\mu^{(n)}(\sigma_n)$$
for all $n$ and any $\sigma_n\in \Omega^G_{V_n}$.

\textbf{Definition 3.} A measure $\mu$ defined by formula
(\ref{rus2.1}) with consistency condition (\ref{rus2.2}) is called a splitting hard core
Gibbs measure with activity $\lambda>0$,
corresponding to the function $z:\,x\in V
\setminus\{x^0\}\mapsto z_x$.

It is known (see Chapter 12, \cite{6}) that any extreme Gibbs measure is splitting Gibbs measure; therefore,  for each given Hamiltonian on Cayley tree, the description of the set of all Gibbs measures is equivalent to the full description of the set of all extreme splitting Gibbs measures.

Let $L(G)$ be the set of edges of a graph $G$. We let $A\equiv
A^G=\big(a_{ij}\big)_{i,j=0,1,2}$ denote the adjacency
matrix of the graph $G$, i.e.,
$$ a_{ij}\equiv a^G_{ij}=\left\{\begin{array}{ll}
1,\ \ \mbox{if}\ \ \{i,j\}\in L(G),\\
0, \ \ \mbox{if} \ \  \{i,j\}\notin L(G).
\end{array}\right.$$

The following theorem presents a condition on $z_x$ ensuring that the measure $\mu^{(n)}$ is consistent.

\textbf{Theorem 1.}\cite{Ro}\label{rust1} \emph{The probability measures
$\mu^{(n)}$, $n=1,2,\ldots$, defined by formula (\ref{rus2.1})
are consistent if and only if the following relations hold for any $x\in V$:}
\begin{equation}\label{rus2.3}\begin{array}{llllll}
z'_{1,x}=\lambda \prod_{y\in S(x)}{a_{10}+
a_{11}z'_{1,y}+a_{12}z'_{2,y}\over
a_{00}+a_{01}z'_{1,y}+a_{02}z'_{2,y}},\\[4mm]
z'_{2,x}=\lambda \prod_{y\in S(x)}{a_{20}+
a_{21}z'_{1,y}+a_{22}z'_{2,y}\over
a_{00}+a_{01}z'_{1,y}+a_{02}z'_{2,y}},
\end{array}
\end{equation}
\emph{where} $z'_{i,x}=\lambda z_{i,x}/z_{0,x}, \ \ i=1,2$.\

In (\ref{rus2.3}), we assume that $z_{0,x} \equiv1$ and $z_{i,x}=z'_{i,x}>0$ for $i=1,2.$ Then
by Theorem 1 there exists a unique $G$-HC Gibbs measure $\mu$ if and only if for any functions
$z:x\in V \longmapsto z_{x}=(z_{1,x},z_{2,x})$ the equality holds:
\begin{equation}\label{ee1}
z_{i,x}=\lambda\prod_{y\in S{(x)}}\frac{a_{i0}+a_{i1}z_{1,y}+a_{i2}z_{2,y}}{a_{00}+a_{01}z_{1,y}+a_{02}z_{2,y}},    i=1,2.
\end{equation}

It is known that we have one-to-one correspondence between the set $V$ of vertices of a Cayley tree
of order $k\geq1$ and the group $G_k$ that is the free product of $k+1$ cyclic groups of second order with the
corresponding generators $a_1,a_2, \ldots, a_{k+1}$ (see \cite{1}).

Let $G_k/\widehat{G}_k=\{H_1,...,H_r\}$ be the quotient group, where
$\widehat{G}_k$ is a normal subgroup of index $r\geq 1.$

\textbf{Definition 4.} The set of vectors $z=\{z_x,x\in G_k\}$
is said to be $\widehat{G}_k$- periodic if  $z_{yx}=z_x$  for all
$\forall x\in G_k, y\in\widehat{G}_k.$\

$G_k$-periodic sets are said to be translation-invariant.

\textbf{Definition 5.} A measure $\mu$ is said to be
$\widehat{G}_k$-periodic if it corresponds to the
$\widehat{G}_k$-periodic set of vectors $z$.

For TIGM in the case $G=\textit{wand}$ the following facts are known:

\begin{itemize}
\item[$\bullet$] In the case $k=2$ ($k=3$) for $\lambda\leq1$ ($\lambda\leq\frac{4}{27}$),
there is a unique TIGM $\nu_0$ and for $\lambda>1$ ($\lambda>\frac{4}{27}$) there are exactly three
TIGMs $\nu_0, \nu_1, \nu_2$ (see \cite{Ro}, \cite{XR1}).

\item[$\bullet$] In the case $k>3$ for $\lambda\leq\lambda_{cr}$, there is a unique TIGM
and for $\lambda>\lambda_{cr}$ there are at least three TIGMs where
$\lambda_{cr}={1\over k-1}\cdot\left({2\over k}\right)^k$ (see \cite{RKh1}).

\item[$\bullet$] In the case $k=2$, the measure $\nu_0$ for $0<\lambda<\lambda_0$ and
the measures $\nu_1, \nu_2$ for $1<\lambda<\lambda_1$ are extreme and the measure $\nu_0$ for $\lambda>\lambda_0$
is not extreme, where $\lambda_0\approx 2.287572$, $\lambda_1\approx 1.303094$ (see \cite{RKh1}).
\end{itemize}

\section{Periodic splitting Gibbs measures in the case $G=\textit{wand}$}\

In the case $G=\textit{wand}$, we write (\ref{ee1}) in the following form:
\begin{equation}\label{ee2}\begin{array}{llllll}
h_{1,x}=\ln\lambda+\sum_{y\in S(x)} \ln{\frac{1+e^{h_{1,y}}}{e^{h_{1,y}}+e^{h_{2,y}}}},\\[4mm]
h_{2,x}=\ln\lambda+\sum_{y\in S(x)} \ln{\frac{1+e^{h_{2,y}}}{e^{h_{1,y}}+e^{h_{2,y}}}},\\
\end{array}
\end{equation}
where $h_{i,x}=\ln z_{i,x}, \ i=1,2$. We study periodic solutions of system (\ref{ee2}).

Let the function $F(\cdot):h=(h_1,h_2)\longmapsto F(h)=(F_1(h),F_2(h))$ be given by

$$F_1(h)=\ln\frac{1+e^{h_1}}{e^{h_1}+e^{h_2}},  \ \     F_2(h)=\ln\frac{1+e^{h_2}}{e^{h_1}+e^{h_2}}.$$

\textbf{Proposition.} \emph{The function $F$ is injective}.

\textbf{Proof.} Necessity. Let $F(h)=F(l)$. Then $F_1(h)=F_1(l)$, $F_2(h)=F_2(l)$,
where $h=(h_1,h_2)$, $l=(l_1,l_2)$. From these equalities we obtain the following system of equations:
\begin{equation}\label{ee29}\begin{array}{llllll}
(1-z_2)(z_1-t_1)+(1+z_1)(z_2-t_2)=0,\\[2mm]
(1+z_2)(z_1-t_1)+(1-z_1)(z_2-t_2)=0.
\end{array}
\end{equation}
Here, $z_i=e^{h_i}, t_i=e^{l_i}, i=1,2.$ It is easy to see that the determinant of system (\ref{ee29}) is nonzero: $\Delta=-2(z_1+z_2)\neq0.$
Therefore, the system (\ref{ee29}) has a unique solution $z_1=t_1$, $z_2=t_2$.

Let $G^{(2)}_k$ be the subgroup of $G_k$ consisting the words of even length.

\textbf{Theorem 2.} \emph{Let $H$ be a normal subgroup of finite index in $G_k$.
Then for HC model each $H$-periodic splitting Gibbs measure is either $G^{(2)}_k$-periodic or translation-invariant.}

\textbf{Proof.} The proof is similar to the proof of Theorem 2 from \cite{MRS} using the result of the Proposition 3.1.

\textbf{Remark 1.} An analogies of Theorem 2 can be proved for a wide class of hard
constraint models.

By Theorem 2, there are only $G^{(2)}_k$-periodic Gibbs measures, and for them from (\ref{ee2}) we obtain the following system of equations:
\begin{equation}\label{ee8} \left\{\begin{array}{ll}
t_{1}=\lambda\left(\frac{1+z_1}{z_1+z_2}\right)^k,\\
t_{2}=\lambda\left(\frac{1+z_2}{z_1+z_2}\right)^k,\\
z_{1}=\lambda\left(\frac{1+t_1}{t_1+t_2}\right)^k,\\
z_{2}=\lambda\left(\frac{1+t_2}{t_1+t_2}\right)^k.\\
\end {array}\right.
\end{equation}
We consider the map $W:R^4\rightarrow R^4$ defined as
\begin{equation}\label{ee4}\left\{\begin{array}{ll}
 t^{'}_1=\lambda\left(\frac{1+z_1}{z_1+z_2}\right)^k,\\
 t^{'}_2=\lambda\left(\frac{1+z_2}{z_1+z_2}\right)^k,\\
 z^{'}_1=\lambda\left(\frac{1+t_1}{t_1+t_2}\right)^k,\\
 z^{'}_2=\lambda\left(\frac{1+t_2}{t_1+t_2}\right)^k.\\
 \end {array}\right.
 \end{equation}
We note that the system (\ref{ee8}) is the equation $z=W(z)$. Therefore, solving the system (\ref{ee8}) is equivalent to finding
fixed points of the map $z^{'}=W(z).$

The next lemma is true.

\textbf{Lemma 1.} \emph{The following sets are invariant under the map $W$:}
$$I_1={\{(t_1,t_2,z_1,z_2)\in R^4:t_1=t_2=z_1=z_2\}},$$
$$I_2={\{(t_1,t_2,z_1,z_2)\in R^4:t_1=t_2,z_1=z_2\}},$$
$$I_3={\{(t_1,t_2,z_1,z_2)\in R^4:t_1=z_1,t_2=z_2\}},$$
$$I_4={\{(t_1,t_2,z_1,z_2)\in R^4:t_1=z_2,t_2=z_1\}}.$$

\textbf{Proof.} The proof is similar to the proof of Lemma 2 from \cite{RK}.

\textbf{Remark 2.} It is difficult to solve system (\ref{ee8}) in the general case, so we will solve it on invariant sets $I_i, \ i=1, 2, 3, 4$. Note that there may be other invariant sets.

\textbf{Theorem 3.} \emph{For HC model in the case $G=\textit{wand}$ the following statements are true:}

\emph{1. For $k\geq2$, $\lambda>0$ on $I_1$ and $I_4$ each $G^{(2)}_k$-periodic splitting Gibbs measure is translation- invariant.
Moreover, this measure coincides with the unique translation-invariant Gibbs measure $\nu_0$.}

\emph{2. For $k\geq2$, $\lambda>0$ on $I_3$ each $G^{(2)}_k$-periodic splitting Gibbs measure is translation-invariant and this measure is not unique.}

\emph{3. Let $k=2$ and $\lambda_{cr}=1$. Then on $I_2$ for $\lambda\geq\lambda_{cr}$ there is exactly one $G^{(2)}_k$-periodic splitting
Gibbs measure which coincides with the unique TIGM $\nu_0$, and for $0<\lambda<\lambda_{cr}$ there are exactly three
$G^{(2)}_k$-periodic splitting Gibbs measures $\nu_0, \mu_1, \mu_2$, where $\mu_1, \mu_2$ are non translation-invariant.}

\emph{4. Let $k=3$ and $\lambda_{cr}=\frac{128}{27}$. Then on $I_2$ for $\lambda\geq\lambda_{cr}$ there is exactly one $G^{(2)}_k$-periodic splitting Gibbs measure which is translation-invariant and for $0<\lambda<\lambda_{cr}$ there are exactly three
$G^{(2)}_k$-periodic splitting Gibbs measures, one of which is translation-invariant and the other two are non translation-invariant.}

\textbf{Proof.} 1. The case $I_1$ is obvious.

\textit{The case $I_4$.} In this case, the system of equations (\ref{ee8}) has the form

\begin{equation}\label{ee9}
\left\{%
\begin{array}{ll}
   z_{1}=\lambda\left(\frac{1+z_{2}}{z_{1}+z_{2}}\right)^{k}, \\[3 mm]
   z_{2}=\lambda\left(\frac{1+z_{1}}{z_{1}+z_{2}}\right)^{k}. \\
\end{array}%
\right.\end{equation}
It suffices to show that the system of functional equations (\ref{ee9}) has only roots of the form
$z_1=z_2$ for any $z_1>0$, $z_2>0$, $\lambda>0$ and $k\geq2$.
Introducing the notation $\sqrt[k]{z_1}=x, \ \sqrt[k]{z_2}=y$, we rewrite the system of equations (\ref{ee9}):
$$
\left\{%
\begin{array}{ll}
   x=\sqrt[k]{\lambda}\left(\frac{1+y^k}{x^k+y^k}\right), \\[3 mm]
   y=\sqrt[k]{\lambda}\left(\frac{1+x^k}{x^k+y^k}\right). \\
\end{array}%
\right.$$
In the last system of equations, subtract the second from the first equation
$$(x-y)(x^k+y^k+\sqrt[k]{\lambda}(x^{k-1}+x^{k-2}y+...+y^{k-1}))=0.$$
Hence, $x=y$, i.e., $(t_1,t_2,z_1,z_2)\in I_1$. So $G^{(2)}_k$-periodic Gibbs measure is translation-invariant and
this measure is unique.

2. \textit{The case $I_3$.} In this case we obtain the system of equations for the TIGM which was studied in \cite{XR1}, \cite{RKh1} and \cite{Ro}.

3. \textit{The case $I_2$ and $k=2$.} In this case we have $z_1=z_2=z$ and $t_1=t_2=t$. Then the system of equations (\ref{ee8}) has the form

\begin{equation}\label{ee100} \left\{\begin{array}{ll}
z=\lambda\left(\frac{1+t}{2t}\right)^{k}, \\[3 mm]
t=\lambda\left(\frac{1+z}{2z}\right)^{k}. \\
\end {array}\right.
\end{equation}
Let $k=2$. Introducing the notation  $\sqrt{z}=x, \ \sqrt{t}=y$ we rewrite the system of equations (\ref{ee100}):
\begin{equation}\label{ee10} \left\{\begin{array}{ll}
x=\sqrt{\lambda}{\frac{1+y^2}{2y^2}}, \\[3 mm]
y=\sqrt{\lambda}{\frac{1+x^2}{2x^2}}. \\
\end {array}\right.
\end{equation}
The system (\ref{ee10}) leads to the following equation
$$\lambda{(1+x^2)^2}-2x{(1+x^2)^2}\sqrt{\lambda}+4x^4=0.$$
We regard the last equation as a quadratic equation for variable $\sqrt{\lambda}=a$
whose solutions have the following forms:
$$a_1=\frac{2x}{1+x^2}, \ a_2=\frac{2x^3}{1+x^2}.$$
Note that for any value $\lambda>0$ the equation
\begin{equation}\label{ee111}
a_2=\sqrt{\lambda}=\frac{2x^3}{1+x^2}
\end{equation}
has a unique solution which corresponds to the unique TIGM for the HC model in the case $G=\textit{wand}$ (see \cite{RKh1} formula (3.15)).

Now from the expression for $a_1=\sqrt{\lambda_1}\equiv\sqrt{\lambda}$ we get the quadratic equation
\begin{equation}\label{ee11}
\sqrt{\lambda}x^2-2x+\sqrt{\lambda}=0,
\end{equation}
Note that the Equation (\ref{ee11}) has solutions for $0<\lambda\leq1$, more exactly,
for $\lambda=1$ it has a unique solution of the form $x_1=x_2=1$ which is also a solution to (\ref{ee111}) for this value of $\lambda$ and for $0<\lambda<1$ it has two positive solutions:
$$x_1=\frac{1+\sqrt{1-\lambda}}{\sqrt{\lambda}}, \ x_2=\frac{\sqrt{\lambda}}{1+\sqrt{1-\lambda}}.$$
From the second equation in (\ref{ee10}), we obtain
$$y_1=\frac{\sqrt{\lambda}}{1+\sqrt{1-\lambda}}, \ y_2=\frac{1+\sqrt{1-\lambda}}{\sqrt{\lambda}}.$$
So for the system of equations (\ref{ee100}) we have solutions of the form $(x_1^2,y_1^2)=(z,t)$ and $(x_2^2,y_2^2)=(t,z)$, where
\begin{equation}\label{ee77}
z=\frac{(1+\sqrt{1-\lambda})^2}{\lambda}, \ t=\frac{\lambda}{(1+\sqrt{1-\lambda})^2}.
\end{equation}
Thus the solutions $(z,t)$ and $(t,z)$ of system of equations (\ref{ee100}) corresponds to the $G^{(2)}_k$-periodic Gibbs measures $\mu_1$ and $\mu_2$ which are different from translation-invariant.

4. \textit{The case $I_2$ and $k=3$.} In this case introducing the notation $\sqrt[3]{z}=x, \ \sqrt[3]{t}=y$ we rewrite the system of equations (\ref{ee100}):
\begin{equation}\label{ee90} \left\{\begin{array}{ll}
x=\sqrt[3]{\lambda}{\frac{1+y^3}{2y^3}}, \\[3 mm]
y=\sqrt[3]{\lambda}{\frac{1+x^3}{2x^3}}. \\
\end {array}\right.
\end{equation}
From the system of equations  (\ref{ee90}) we have

\begin{equation}\label{ee91} \left\{\begin{array}{ll}
x=f(y), \\[3 mm]
y=f(x), \\
\end {array}\right.
\end{equation}
where
$$f(x)=\sqrt[3]{\lambda}\frac{1+x^3}{2x^3}.$$
It is easy to check that the equation $f(x)=x$ has a unique positive solution for any $\lambda>0$, which corresponds to the unique TIGM.

Moreover, roots of the equation $f(x)=x$ are clearly roots of the equation $f(f(x))=x.$
To find the roots of the equation $f(f(x))=x$ that differ from roots of $f(x)=x$, we must therefore consider the equation
\begin{equation}\label{eee1}
\frac{x-f(f(x))}{x-f(x)}=0.
\end{equation}
The equation (\ref{eee1}) is equivalent to the equation
$$b^2(1+x^3)^2-2bx(1+x^3)-4x^5=0$$
where $b=\sqrt[3]{\lambda}$. We consider the last equation as a quadratic equation for $b$.
It has one positive solution:
$$b=\frac{x+x\sqrt{1+4x^3}}{1+x^3}>0.$$
We consider next equation:
\begin{equation}\label{ee92}
b=\sqrt[3]{\lambda}=\frac{x+x\sqrt{1+4x^3}}{1+x^3}=\varphi(x).
\end{equation}

It is easy to see that the function $\varphi(x)$ increases for $0<x\leq\sqrt[3]{2}$
and decreases for $x\geq\sqrt[3]{2}$, i.e., $x_{max}=\sqrt[3]{2}$ and
$\varphi(x_{max})=\frac{4\sqrt[3]{2}}{3}=b=\sqrt[3]{\lambda_{cr}}$ (see Fig.1, a)).
\begin{center}
\includegraphics[width=6cm]{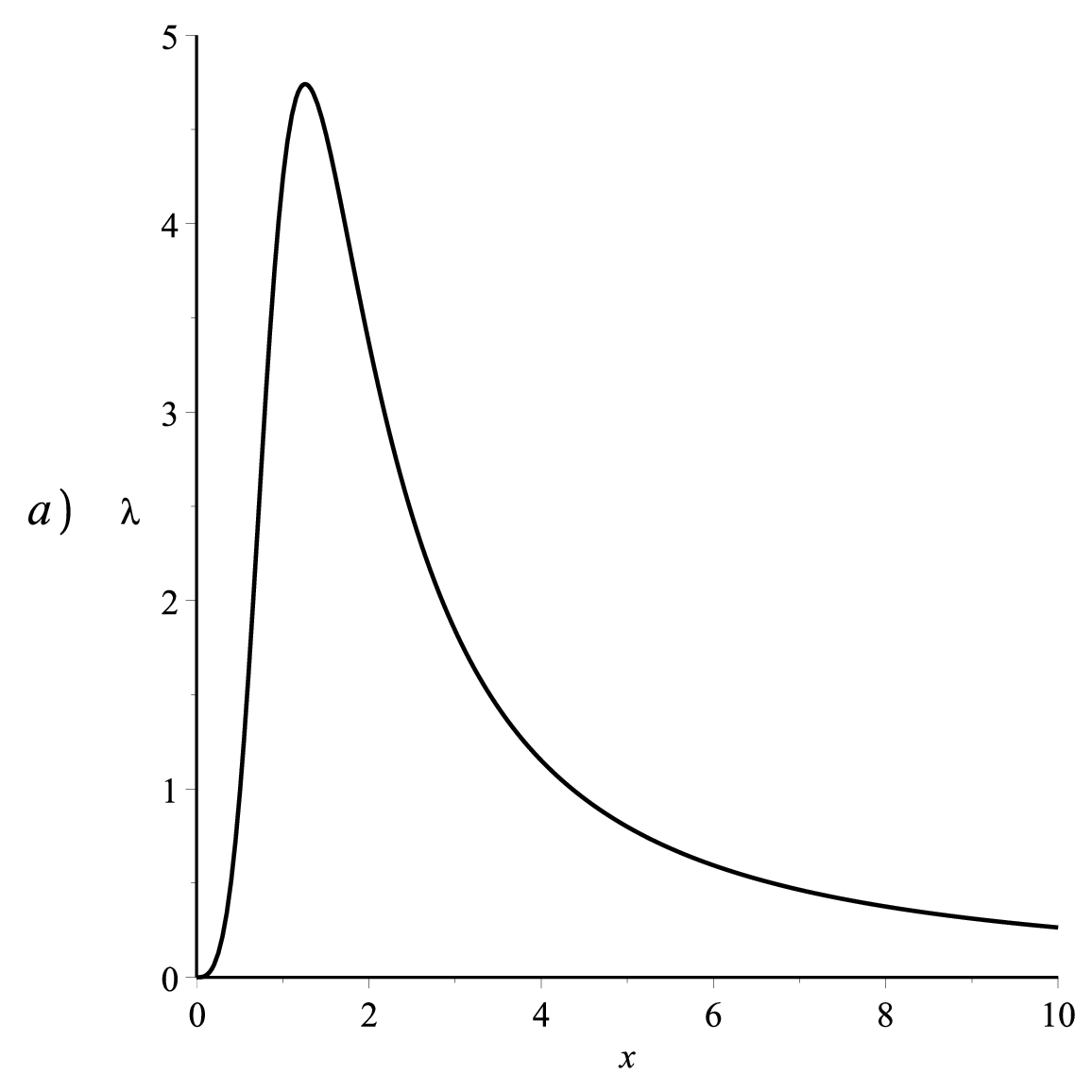} \  \includegraphics[width=6cm]{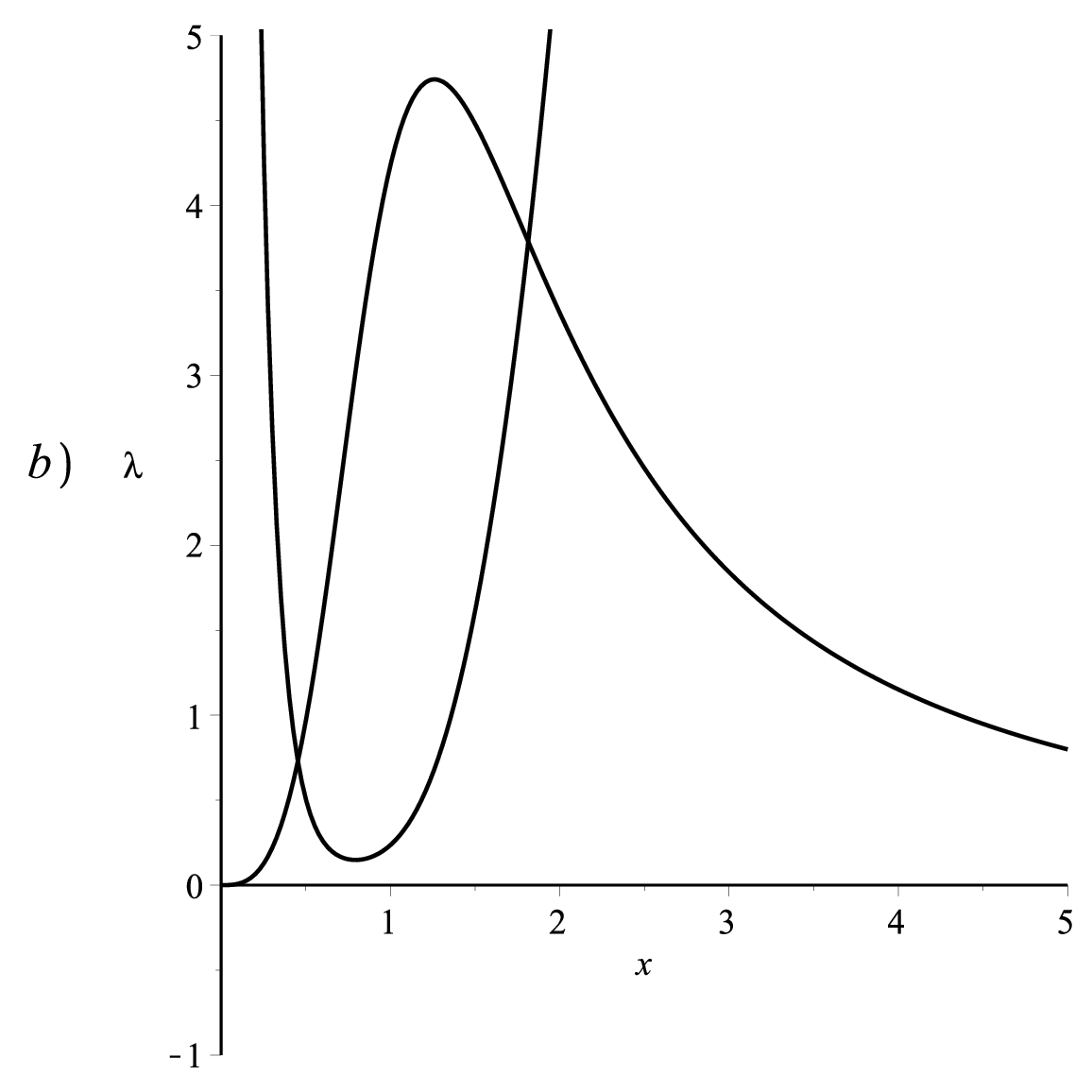}
\end{center}
\begin{center}{\footnotesize \noindent
 Fig.~1. a) Graph of the function $\varphi^3(x)$. b) Graph of the function $\varphi^3(x)$ (upper curve) and graph of the function $\psi^3(x)$ (lower curve).}
\end{center}

Thus, the equation $\lambda=\varphi^3(x)$ has two solutions if $0<\lambda<\lambda_{cr}=\frac{128}{27}$, one solution if $\lambda=\lambda_{cr}$, and no solution if $\lambda>\lambda_{cr}$.

Note that if $\lambda=\lambda_{cr}=\frac{128}{27}$ then solution (\ref{ee90}) has the form: $(\sqrt[3]{2}, \sqrt[3]{2})$,
i.e., this solution corresponds to the TIGM which exists for any $\lambda>0$, and measures corresponding to the two existing
solutions for $0<\lambda<\lambda_{cr}$ are $G^{(2)}_k$-periodic different from translation-invariant.

\textbf{Remark 3.} In \cite{XR1}, TIGMs was investigated for the HC model in the case $G=\textit{wand}$
and a similar method was applied which was used in the proof of part 4 of Theorem 3.6.
For TIGMs, the equation (\ref{ee92}) is given by
\begin{equation}\label{ee96}
\sqrt[3]{\lambda}=\frac{x^2(x^3-3)+\sqrt{x^4(x^3+3)^2+4x}}{2x(x^3+1)}=\psi(x).
\end{equation}
The equation shows that the functions $\varphi^3(x)$ and $\psi^3(x)$ differ and they intersect at two points: $x_1\approx0.4531316267$,
$x_2\approx1.813976199$, i.e., measures corresponding to these solutions are TIGMs (see Fig.1, b)).

\textit{The case $I_2$ and $k\geq2$.} In this case we rewrite the system of equations
(\ref{ee100}):
\begin{equation}\label{t1} \left\{\begin{array}{ll}
z=h(t), \\[3 mm]
t=h(z), \\
\end {array}\right.
\end{equation}
where $h(x)=\alpha\left(1+\frac{1}{x}\right)^{k}, \alpha=\frac{\lambda}{2^k}$.

The following lemma is known.

\textbf{Lemma 2.} \cite{K} \textit{Let $f:[0,1]\rightarrow [0,1]$
be a continuous function with a fixed point $\xi \in (0,1)$. We
assume that $f$ is differentiable at $\xi$ and $f^{'}(\xi)<-1.$
Then there exist points $x_0$ and $x_1$, $ 0\leq x_0<\xi<x_1
\leq1,$ such that $f(x_0)=x_1$ and $f(x_1)=x_0.$}\

\textbf{Theorem 4.} \emph{Let $k\geq 2$ and $\lambda_{cr}=2^k(k-1)\left({k-1\over k}\right)^k.$ Then
for HC model in the case $G=\textit{wand}$ on $I_2$ for $0<\lambda<\lambda_{cr}$ there exist at least three $G^{(2)}_k$-periodic splitting Gibbs measures, one of which is translation-invariant and the other two are non translation-invariant.}

\textbf{Proof.} Since $h(x)>\alpha$, we have  $z>\alpha$ and $t>\alpha$.
The function $h(x)$ is decreasing and
$h_{max}=h(\alpha)=\alpha\left(1+\frac{1}{\alpha}\right)^{k}=\beta.$
Moreover, the function $h(x)$ is a continuous and
differentiable in $[\alpha, \beta]$. It follows from the above argument that the
equation $h(x)=x$ has a unique solution $x=\xi.$

We rewrite the equality $h(\xi)=\xi$ as follows
$$\alpha\left(1+\frac{1}{\xi}\right)^{k-1}=\frac{\xi^2}{1+\xi}.$$
Using the last equality, from the inequality
$$h'(\xi)=-\frac{\alpha k}{\xi^2}\left(1+\frac{1}{\xi}\right)^{k-1}=-\frac{k}{1+\xi}<-1$$
we get $\xi<k-1.$

Since $\xi\in(\alpha, \beta)$ is a fixed point of the function $h$, we have
$$\lambda=2^k\xi\left(\frac{\xi}{1+\xi}\right)^{k}=\phi(\xi).$$
Note that $\phi'(\xi)>0,$ i.e., the function $\phi(\xi)$ is increasing. Hence, for $\xi<k-1$ we have $\phi(\xi)<\phi(k-1)$, i.e.
$$\lambda_{max}=\lambda_{\rm cr}=2^k(k-1)\left({k-1\over k}\right)^k.$$
Consequently, by Lemma 3.6 if $\lambda<\lambda_{\rm cr}$ then the system
(\ref{t1}) has three solutions $(\xi,\xi), \ (z_0,t_0)$ and
$(t_0,z_0)$.

\section{Extremality of periodic splitting Gibbs measures in the case $G=\textit{wand}$}\

We have $G^{(2)}_k$-periodic splitting Gibbs measures $\mu_1$ and $\mu_2$ for $k=2$.
To study their (non) extremality we use the methods from  \cite{KS}, \cite{KR}, \cite{Kr1} and \cite{MSW} for TIGM.
For each translation-invariant measure we consider a tree-indexed Markov chain with states $\{0,1,2\}$, i.e.,
suppose we are given a Cayley tree with set vertices $V$, a
probability measure $\nu$, and a probability transition matrix
$\mathbb P = \left(P_{ij}\right)$ on $\{0,1,2\}$.
Using transition probabilities given the value of its parent, regardless of everything else we can construct
a tree - indexed by a Markov chain $X: V \to \{0,1,2\}$ by choosing $X(x^0)$ according to $\nu$ and choosing $X(v)$,
for each vertex $v\ne x^0$.

Since translation-invariant measures are obtained for $(t_1,t_2)=(z_1,z_2)$, matrix $\mathbb P$ depends only on
$z_1$($=t_1$) and $z_2$($=t_2$), more precisely,
$$
\mathbb P=\left(%
\begin{array}{cccccc}
 0 &  {z_1\over z_1+z_2} &  {z_2\over z_1+z_2} \\[3 mm]
  {1\over 1+ z_1} & { z_1\over 1+ z_1} & 0 \\[3 mm]
  {1\over 1+ z_2} & 0 & { z_2\over 1+ z_2} \\
  \end{array}%
\right)
$$
But, in the case of periodic measures, the matrix $\mathbb P$ depends on $t_1$, $t_2$, $z_1$ and $z_2$, where $t_1\ne z_1$, $t_2\ne z_2$ and $(t_1, t_2, z_1, z_2)$ are solutions of the system of equations (\ref{ee8}). We consider measures $\mu_1$ and $\mu_2$ corresponding to the set of $I_2: t_1=t_2=t, z_1=z_2=z$. In addition, note that $zt=1$. Then the transitions probabilities matrix $P_{il}$ defined by the given periodic Gibbs measure $\mu_1$ (resp. $\mu_2$) $\mathbb P\equiv \mathbb P_{z, t}=\mathbf{P_{\mu_1}}$ (resp. $\mathbb P\equiv \mathbb P_{t, z}=\mathbf{P_{\mu_2}}$) is the product of two transition probabilities matrices:
$$
\mathbf{P_{\mu_1}}=\mathbb P_{z}\mathbb P_{t}=\begin{pmatrix}
0 & \frac{1}{2} & \frac{1}{2}\\ \frac{1}{1+z} & \frac{z}{1+z} & 0 \\ \frac{1}{1+z} & 0 & \frac{z}{1+z}\end{pmatrix}\cdot
\begin{pmatrix}0 & \frac{1}{2} & \frac{1}{2}\\ \frac{1}{1+t} & \frac{t}{1+t} & 0 \\ \frac{1}{1+t} & 0 & \frac{t}{1+t}\end{pmatrix}=
$$

\begin{equation}\label{rus2.7}=\begin{pmatrix}
\frac{1}{1+t} & \frac{t}{2(1+t)} & \frac{t}{2(1+t)}
\\\frac{z}{(1+z)(1+t)} & \frac{t+3}{2(1+z)(1+t)} & \frac{1}{2(1+z)} \\ \frac{z}{(1+z)(1+t)} & \frac{1}{2(1+z)} & \frac{t+3}{2(1+z)(1+t)}\end{pmatrix}
\end{equation}

Thus, the matrix $\mathbf{P_{\mu_1}}$ defines a Markov chain on the Cayley tree of order $k^2$, which consists of
the vertices of the tree $\Gamma^k$ in even places.

So a sufficient condition  (i.e., the Kesten-Stigum condition, see \cite{KS}) for non-extremality of a Gibbs measure $\mu_1$ corresponding to the matrix $\mathbf{P_{\mu_1}}$ is that $k^2s_2^2>1$, where $s_2$ is the second largest (in absolute value) eigenvalue of $\mathbf{P_{\mu_1}}$.

It is clear that the eigenvalues of this matrix are
$$s_1=1, \ s_2=s_3=\frac{1}{z+t+2}.$$
We have solutions of the form (\ref{ee77}) for $k=2$. By virtue of the symmetry of the solutions, the region of non-extremality
of the measure $\mu_2$ coincides with the region of non-extremality of the measure $\mu_1$. Therefore, it is sufficient
to check the condition of non-extremality of the measure $\mu_1$ for $k=2$. For this, we calculate $z+t$:
$$z+t=\frac{2(2-\lambda)}{\lambda}.$$
Then from $4s_2^2>1$ we obtain $\lambda>2$, but measures $\mu_1$ and $\mu_2$ exist for $0<\lambda<1$.
Hence, this measures should be extreme, which we shall check below.

Let us first give some necessary definitions from \cite{MSW}.
If from a Cayley tree $\Gamma^k$ we remove an arbitrary edge $\langle x^0, x^1\rangle=l\in L$, then it is divided into two
components $\Gamma^k_{x^0}$ and $\Gamma^k_{x^1}$, each called semi-infinite Cayley tree or Cayley subtree.

We consider the finite complete subtrees $\mathcal T$, that are the initial points of Cayley tree $\Gamma^k_{x^0}$. The boundary
$\partial \mathcal T$ of the subtree $\mathcal T$ consists of the neighbors which are on $\Gamma^k_{x^0}\setminus \mathcal T$.
We identify the subgraphs of $\mathcal T$ with their vertex sets and write $E(A)$ for the edges within
a subset $A$ and $\partial A$ for the boundary of $A$.

In \cite{MSW}, the key ingredients are two quantities, $\kappa$ and $\gamma$. Both are properties of the
collection of Gibbs measures $\{\mu^\tau_{{\mathcal T}}\}$, where the boundary condition $\tau$ is fixed and $\mathcal T$ ranges
over all initial finite complete subtrees of $\Gamma^k_{x^0}$. For a given subtree $\mathcal T$ of $\Gamma^k_{x^0}$ and a vertex $x\in\mathcal T$, we write $\mathcal T_x$ for the half tree growing from root $x$. When $x$ is not the root of $\mathcal T$, let $\mu_{\mathcal T_x}^s$ denote the (finite-volume) Gibbs measure in which the parent of $x$ has its spin fixed to $s$ and the configuration on the bottom boundary ${\mathcal T}_x$ (i.e. on $\partial {\mathcal T}_x\setminus \{\mbox{parent of}\ \ x\}$) is specified by $\tau$.

For two measures $\mu_1$ and $\mu_2$ on $\Omega$, $\|\mu_1-\mu_2\|_x$ denotes the variation distance between
the projections of $\mu_1$ and $\mu_2$ onto the spin at $x$, i.e.,
$$\|\mu_1-\mu_2\|_x={1\over 2}\sum_{i=0}^2|\mu_1(\sigma(x)=i)-\mu_2(\sigma(x)=i)|.$$
Let $\eta^{x,s}$ be the configuration $\eta$ with the spin at $x$ set to $s$.

Following \cite{MSW} define
$$\kappa\equiv \kappa(\mu)={1\over2}\max_{i,j}\sum_{l=0}^2|P_{il}-P_{jl}|;$$
$$\gamma\equiv\gamma(\mu)=\sup_{A\subset \Gamma^k}\max\|\mu^{\eta^{y,s}}_A-\mu^{\eta^{y,s'}}_A\|_x,$$
where the maximum is taken over all boundary conditions $\eta$, all sites $y\in \partial A$, all neighbors $x\in A$ of $y$, and all spins $s, s'\in \{0,1,2\}$.

It is known that a sufficient condition for extremality of the translation-invariant Gibbs measure $\mu$ is that $k\kappa(\mu)\gamma(\mu)<1$,
but for the considered $G^{(2)}_k$-periodic measures $\mu_i, i=1,2$ this condition is $k^2\kappa(\mu_i)\gamma(\mu_i)<1$.

Using (\ref{rus2.7}) we obtain
$$\kappa=\frac{z}{(z+1)^2}.$$
By virtue of the symmetry of the solutions, the region of extremality of the measure $\mu_2$ coincides with the region of extremality of the measure $\mu_1$. It is known from \cite{RKh1} that $\gamma=\kappa$. Hence, for extremality of the measure $\mu_1$ (also for an measure $\mu_2$) we obtain the inequality
$$k^2\kappa(\mu_1)\gamma(\mu_1)=\frac{4z^2}{(z+1)^4}<1$$
which is true for any values $z$ in particular, for a solution of the form (\ref{ee77}) which exists for $0<\lambda<1$.
Consequently, in the case $k=2$, the condition of the extremality measures $\mu_1$ and $\mu_2$ is satisfied for any values $0<\lambda<1$, i.e.,
in an area of the existence of these measures.

So, we have proved the following theorem.

\textbf{Theorem 5.} \emph{Let $k=2$. Then for the HC model in the case $G=\textit{wand}$ $G^{(2)}_k$-periodic splitting Gibbs measures
$\mu_1$ and $\mu_2$ are extreme for $0<\lambda<1$.}

\textbf{Remark 4.} Since in the case $k=3$ we do not have an explicit form of the solution of the system of equations (\ref{ee90}), it is very difficult to investigate (non) extremality of the corresponding periodic Gibbs measures. Therefore, this question is still open.

\end{document}